\documentclass[]{elsart3}
\usepackage{times}
\usepackage{amssymb}
\usepackage{amsfonts}
\usepackage{eucal}
\usepackage{revsymb}
\usepackage{graphicx}
\usepackage{epsfig}

\begin{document}

\begin{frontmatter}
For Computer Physics Communications: Proceedings of CCP$\;$2004
\title{ Fluid Coexistence close to Criticality:\\ Scaling Algorithms for Precise Simulation}

\author{Young C.\ Kim and Michael E.\ Fisher\corauthref{cor1}}
\corauth[cor1]{Corresponding Author: xpectnil@ipst.umd.edu}
\address{Institute for Physical Science and Technology, University of Maryland, College Park, Maryland 20742 USA}

\begin{abstract}

A novel algorithm is presented that yields precise estimates of coexisting liquid and gas densities, $\rho^{\pm}(T)$, from grand canonical Monte Carlo simulations of model fluids near criticality. The algorithm utilizes data for the isothermal minima of the moment ratio $Q_{L}(T;\langle\rho\rangle_{L})$ $\equiv\langle m^{2}\rangle_{L}^{2}/\langle m^{4}\rangle_{L}$ in $L$$\,\times$$\, \cdots$$\,\times$$\, L$ boxes, where $m=\rho-\langle\rho\rangle_{L}$. When $L$$\,\rightarrow$$\,\infty$ the minima, $Q_{\mbox{\scriptsize m}}^{\pm}(T;L)$, tend to zero while their locations, $\rho_{\mbox{\scriptsize m}}^{\pm}(T;L)$, approach $\rho^{+}(T)$ and $\rho^{-}(T)$. Finite-size scaling relates the ratio {\boldmath $\mathcal Y$}$\,=\,$$(\rho_{\mbox{\scriptsize m}}^{+}-\rho_{\mbox{\scriptsize m}}^{-})/\Delta\rho_{\infty}(T)$ {\em universally} to $\frac{1}{2}(Q_{\mbox{\scriptsize m}}^{+}+Q_{\mbox{\scriptsize m}}^{-})$, where $\Delta\rho_{\infty}$$\,=\,$$\rho^{+}(T)-\rho^{-}(T)$ is the desired width of the coexistence curve. Utilizing the exact limiting $(L$$\,\rightarrow\,$$\infty)$ form, the corresponding scaling function can be generated in recursive steps by fitting overlapping data for three or more box sizes, $L_{1}$, $L_{2}$, $\cdots$, $L_{n}$. Starting at a $T_{0}$ sufficiently far below $T_{\mbox{\scriptsize c}}$ and suitably choosing intervals $\Delta T_{j}$$\,=\,$$T_{j+1}-T_{j}$$\,>\,$$0$ yields $\Delta\rho_{\infty}(T_{j})$ and precisely locates $T_{\mbox{\scriptsize c}}$. 

The algorithm has been applied to simulation data for a hard-core square-well fluid and the restricted primitive model electrolyte for sizes up to $L/a = 8$-$12$ (where $a$ is the hard-core diameter): the coexistence curves can be computed to a precision of $\pm 1$-$2\%$ of $\rho_{\mbox{\scriptsize c}}$ up to $|T-T_{\mbox{\scriptsize c}}|/T_{\mbox{\scriptsize c}}=10^{-4}$ and $10^{-3}$, respectively. Universality of the scaling functions and the exponent $\beta$ is verified and the $(T_{\mbox{\scriptsize c}},\rho_{\mbox{\scriptsize c}})$ estimates confirm previous values based on data from above $T_{\mbox{\scriptsize c}}$. The algorithm extends directly to calculating the diameter, $\rho_{\mbox{\scriptsize diam}}(T) \equiv \frac{1}{2} (\rho^{+} + \rho^{-})$, and can lead to estimates of the Yang-Yang ratio. Furthermore, a new, explicit approximant for the basic scaling function {\boldmath $\mathcal Y$} permits straightforward estimates of $\Delta\rho_{\infty}(T)$ from limited $Q$-data when Ising-type criticality may be assumed.
\end{abstract}


\end{frontmatter}

\section{Introduction}
\label{sec1}

In recent years computer simulation has been an important tool to understand the critical behavior of fluids \cite{pan}. Various programming algorithms and techniques have been developed to enhance calculations with large-scale computers. However, determining phase boundaries, critical points and the universality classes of complex fluids, such as electrolytes, polymers, colloids, etc., has been and still is a great challenge. Here we present in detail, together with a new, `economical' extension, a powerful method developed recently \cite{kim:fis:lui} to estimate precisely coexisting liquid and gas densities, $\rho^{+}(T)$ and $\rho^{-}(T)$, very close to the critical temperature, $T_{\mbox{\scriptsize c}}$. Precise values of $\rho^{\pm}(T)$ can then provide critical parameters and reveal critical exponents, via
 \begin{equation}
  \Delta\rho_{\infty}(T)\equiv \rho^{+}-\rho^{-} \approx B|t|^{\beta}, \hspace{0.05in} t \equiv (T-T_{\mbox{\scriptsize c}})/T_{\mbox{\scriptsize c}}.   \label{eq0}
 \end{equation}

To determine $\rho^{\pm}(T)$ in simulations it has been customary to observe the grand canonical equilibrium distribution function, ${\mathcal P}_{L}(\rho;T)$, of the density, $\rho\equiv N/V$, at constant $T<T_{\mbox{\scriptsize c}}$, where $N$ and $V\equiv L^{d}$ are the particle number and volume of the system, respectively. For large $L$ below $T_{\mbox{\scriptsize c}}$, the distribution ${\mathcal P}_{L}(\rho;T)$ exhibits two well separated peaks located near $\rho^{\pm}(T)$. Examining these peaks with aid of the equal-weight prescription \cite{bor:kap} provides reasonable estimates for $\rho^{\pm}(T)$. However, when $T$ approaches $T_{\mbox{\scriptsize c}}$, the peaks broaden, overlap strongly, and can no longer be separated uniquely thereby precluding reliable estimation of $\rho^{\pm}(T)$ \cite{ork:fis:pan}. As well known, the underlying reason is the divergence of the bulk correlation length at criticality as $|t|^{-\nu}$.

To obtain better estimates of $\rho^{\pm}(T)$ valid closer to $T_{\mbox{\scriptsize c}}$, we examine instead the $Q$-parameter defined by \cite{bin}
 \begin{equation}
   Q_{L}(T;\langle\rho\rangle_L ) \equiv \langle m^{2}\rangle_{L}^{2}/\langle m^{4}\rangle_{L}, \hspace{0.1in} m = \rho-\langle\rho\rangle_{L},   \label{eq1}
 \end{equation}
where $\langle\cdot\rangle_{L}$ denotes a finite-size grand canonical expectation value at fixed $T$ and chemical potential, $\mu$, in a cubic box of dimensions $L$$\,\times\,$$L$$\,\times\,$$\cdots$$\,\times\,$$L$ with periodic boundary conditions. Below $T_{\mbox{\scriptsize c}}$ one finds that $Q_{L}(T;\langle\rho\rangle_L )$ exhibits two minima, say $Q_{\mbox{\scriptsize m}}^{\pm}(T;L)$, at densities $\rho_{\mbox{\scriptsize m}}^{\pm}(T;L)$ near $\rho^{\pm}(T)$ \cite{bin:lan}: see Fig.\ 1.
\begin{figure}[ht]
\vspace{-0.9in}
\centerline{\epsfig{figure=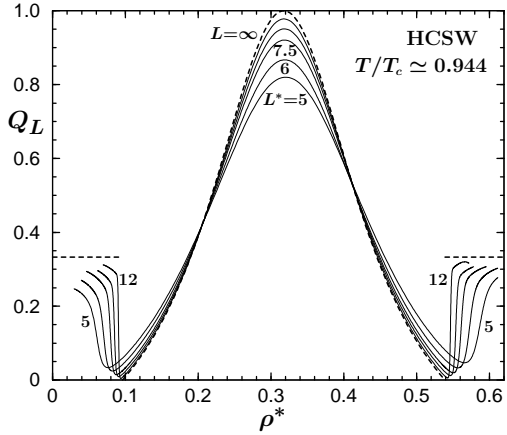,width=3.2in,angle=0}}
\vspace{-1.1in}
\caption{Plots of simulation data for $Q_{L}(T;\langle\rho\rangle_{L})$ vs.\ $\rho^{\ast}$$\equiv$$\langle\rho\rangle a^{3}$ for the HCSW fluid at $T/T_{\mbox{\scriptsize c}}\simeq 0.944$ showing minima that approach the limiting coexistence values $\rho^{+}(T)$ and $\rho^{-}(T)$. The solid curves are for $L^{\ast}\equiv L/a=12,9,7.5,6$ and $5$ (where $a$ is the hard-core diameter); the dashed lines represent the {\em exact} limiting form (for the estimated values of $\rho^{+}$ and $\rho^{-}$). \label{fig1}}
\end{figure}
 When $L$$\,\rightarrow\,$$\infty$ the heights, $Q_{\mbox{\scriptsize m}}^{\pm}(T;L)$, of these minima decay to zero while their locations, $\rho_{\mbox{\scriptsize m}}^{\pm}(T;L)$, approach the desired coexistence values $\rho^{\pm}(T)$. Thus understanding the behavior of the minima is potentially rewarding.

To that end, this article explains an {\em unbiased} scaling algorithm which utilizes calculated values of $Q_{\mbox{\scriptsize m}}^{\pm}(T;L)$ and $\rho_{\mbox{\scriptsize m}}^{\pm}(T;L)$ to obtain precise estimates of the coexistence-curve width or density discontinuity, $\Delta\rho_{\infty}(T)$. By ``unbiased'' we specifically mean that not only are the critical parameters $T_{\mbox{\scriptsize c}}$ and $\rho_{\mbox{\scriptsize c}}$ left open but, also, {\em no assumptions} regarding the value of the exponent $\beta$ in (\ref{eq0}) {\em or} regarding the universality class of the critical point are made (in contrast to earlier approaches \cite{bru:wil,kim:fis}). The algorithm has been applied to a hard-core square-well (HCSW) fluid and to the restricted primitive model (RPM) electrolyte. Although not demonstrated here, the algorithm extends directly \cite{kim:fis:lui,kim:fis2} to accurately estimate the diameter,
 \begin{equation}
 \rho_{\mbox{\scriptsize diam}}(T)\equiv\mbox{$\frac{1}{2}$}[\rho^+(T) + \rho^-(T)].   \label{eq.diam}
 \end{equation}
Furthermore, studying $(Q_{\mbox{\scriptsize m}}^{+}-Q_{\mbox{\scriptsize m}}^{-})$ provides an effective way \cite{kim:fis:lui,kim:fis2} of estimating the strength, ${\mathcal R}_{\mu}$, of the Yang-Yang anomaly, namely, the relative divergence at criticality of the second derivative of the chemical potential on the phase boundary, $d^{\,2}\mu_{\sigma}/dT^{2}$ \cite{fis:ork}.

Finally, on the basis of an explicit expression, approximate but accurate, for a crucial scaling function relating $\Delta\rho_{\infty}(T)$ to the difference $\rho_{\mbox{\scriptsize m}}^{+}(T)-\rho_{\mbox{\scriptsize m}}^{-}(T)$, we demonstrate a simple, albeit biased algorithm that requires only limited data for the $Q$-minima: this should be valuable when, as usual, it may be safely assumed that the criticality is of Ising character \cite{bru:wil,kim:fis}.

\section{Theoretical Background}
\label{sec2}

For sufficiently large $L$ at fixed $T$$\,<\,$$T_{\mbox{\scriptsize c}}$ it is well established \cite{bor:kap,bin:lan} that the density distribution, ${\mathcal P}_{L}(\rho;T)$, asymptotically approaches a sum of two Gaussian peaks which can be written as
 \begin{eqnarray}
   {\mathcal P}_{L}(\rho;T) & \approx & C_{L}\left\{ \chi_{-}^{-1/2}\exp[-\beta(\rho-\rho^{-})^{2}L^{d}/2\chi_{-}] \right. \nonumber \\ 
  &  & \left. \hspace{0.2in} + \chi_{+}^{-1/2}\exp[-\beta(\rho-\rho^{+})^{2}L^{d}/2\chi_{+}]\right\} \nonumber \\
  &  & \hspace{0.2in}\times \exp[\beta\rho(\mu-\mu_{\sigma})L^{d}], \label{eq2}
 \end{eqnarray}
where $\beta$$\,=\,$$1/k_{\mbox{\scriptsize B}}T$ while $\chi_{\pm}(T)$ are the infinite-volume susceptibilities [defined via $\chi$$\,=\,$$(\partial\rho/\partial\mu)_{T}$] evaluated at $\rho$$\,=\,$$\rho^{\pm}(T)\pm$, and $C_{L}(\mu,T)$ is a normalization factor. From this expression the parameter $Q_{L}(T;\langle\rho\rangle_L )$ can be calculated readily and, in particular, the limiting form $Q_{\infty}(T;\langle\rho\rangle_{\infty})$, shown by the dashed lines in Fig.\ 1, can be derived. However, when criticality is approached at fixed $L$, the two-Gaussian representation of ${\mathcal P}_{L}(\rho;T)$ becomes less accurate and it fails badly near criticality. 

In the critical region, on the other hand, the behavior of $Q_{L}(T;\langle\rho\rangle_{L})$ can be understood via finite-size scaling theory\cite{fis:bar}, recently extended to incorporate pressure-mixing in the scaling fields $\tilde{t}$ and $\tilde{\mu}$ \cite{kim:fis3} (which is essential for describing the Yang-Yang anomaly \cite{fis:ork}). For the $Q$-parameter, which depends on the three variables $L$, $T$, and $\langle\rho\rangle_{L}$, finite-size scaling provides the asymptotic, $t$$\,\rightarrow\,$$0$, reduced, two-variable representation
 \begin{equation}
  Q_{L}(T;\langle\rho\rangle_{L}) \approx \mbox{\boldmath $\mathcal Q$}(tL^{1/\nu};\; \Delta\rho/|t|^{\beta}),
 \end{equation}
where {\boldmath $\mathcal Q$}$(x,y)$ is the scaling function, while $\Delta\rho = \langle\rho\rangle_{L}-\rho_{\mbox{\scriptsize c}}$ and, as above, $\nu$ is the correlation length exponent.

It then follows that the minima, $Q_{\mbox{\scriptsize m}}^{+}(T;L)$ and $Q_{\mbox{\scriptsize m}}^{-}(T;L)$, and their corresponding displacements, $[\rho_{\mbox{\scriptsize m}}^{+}(T;L)-\rho_{\mbox{\scriptsize c}}]$ and $[\rho_{\mbox{\scriptsize c}}-\rho_{\mbox{\scriptsize m}}^{-}(T;L)]$, should, on approach to criticality, all reduce to functions of the scaled variable $x$$\,=\,$$tL^{1/\nu}$ alone. Accordingly, the average of $Q_{\mbox{\scriptsize m}}^{+}$ and $Q_{\mbox{\scriptsize m}}^{-}$ and, using (\ref{eq0}) and (\ref{eq.diam}), the normalized density deviation\footnote{The definition of $y$ adopted here differs from that used in \cite{kim:fis:lui} by a factor $\frac{1}{2}$.}
 \begin{equation}
  y \equiv [\langle\rho\rangle_{L}-\rho_{\mbox{\scriptsize diam}}(T)]/\Delta\rho_{\infty}(T), \label{eq.y}
 \end{equation}
should scale likewise. Thus we may anticipate (but should plan to {\em check} in applications) the scaling expressions
 \begin{eqnarray}
  \bar{Q}_{\mbox{\scriptsize min}}(T;L)&\equiv&\mbox{$\frac{1}{2}$}(Q_{\mbox{\scriptsize m}}^{+}+Q_{\mbox{\scriptsize m}}^{-}) \approx {\mathcal M}(tL^{1/\nu}), \label{eq3}  \\
  \Delta y_{\mbox{\scriptsize min}}(T;L) &\equiv& (y_{\mbox{\scriptsize m}}^{+}-y_{\mbox{\scriptsize m}}^{-}) = \frac{\rho_{\mbox{\scriptsize m}}^{+}-\rho_{\mbox{\scriptsize m}}^{-}}{\Delta\rho_{\infty}(T)} \nonumber \\
  &\approx& {\mathcal N}(tL^{1/\nu}), \label{eq4}
 \end{eqnarray}
where ${\mathcal M}(\cdot)$ and ${\mathcal N}(\cdot)$ are appropriate scaling functions, which when properly normalized should be universal.

Before proceeding further, notice that the normalizing divisor $\Delta\rho_{\infty}(T)$ in (\ref{eq.y}) and (\ref{eq4}) is just the true width of the coexistence curve that we wish to estimate!

Now, at least in principle, one may eliminate the scaling variable $x=tL^{1/\nu}$ between (\ref{eq3}) and (\ref{eq4}), e.g., by solving (\ref{eq3}) for $x$ and substituting in (\ref{eq4}), to obtain $\Delta y_{\mbox{\scriptsize min}}$ as a universal function of $\bar{Q}_{\mbox{\scriptsize min}}$, say, {\boldmath $\mathcal Y$}$(q)$. Of course, this function is not known {\em a priori}. However, the two-Gaussian limiting form (\ref{eq2}) for the density distribution ${\mathcal P}_{L}(\rho;T)$, which is easily seen to obey scaling close to $T_{\mbox{\scriptsize c}}$ when $x$ is large, can be used straightforwardly \cite{kim:fis3} to study the minima of $Q_{L}(T;\langle\rho\rangle_{L})$. In this limit we thence obtain the {\em exact} and {\em universal} expansion
 \begin{equation}
  \Delta y_{\mbox{\scriptsize min}} = 1 + \mbox{$\frac{1}{2}$}q + O(q^{2}), \label{eq5}
 \end{equation}
in terms of the auxiliary variable
 \begin{equation}
  q\equiv \bar{Q}_{\mbox{\scriptsize min}}\ln (4/e\bar{Q}_{\mbox{\scriptsize min}}). \label{eq10}
 \end{equation}
As we will explain, this result provides a route to the recursive, numerical construction of the full universal scaling function {\boldmath $\mathcal Y$}$(q)$ and, furthermore, to the evaluation of $\Delta\rho_{\infty}(T)$ and $T_{\mbox{\scriptsize c}}$.

\section{Scaling Algorithm}
\label{sec3}

The basic idea of our algorithm is to fit data for the $Q$-minima to the formula (\ref{eq5}), starting at some temperature $T_{0}$ far enough below $T_{\mbox{\scriptsize c}}$ that the two-Gaussian form (\ref{eq2}) is a good approximation, and then to extend the fits progressively to higher temperatures checking consistency with scaling, i.e., the uniqueness of {\boldmath $\mathcal Y$}$(q)$, as the calculations proceed. To make the fits, it is just the sought-for values of $\Delta\rho_{\infty}(T)$ that must be selected: and in order to vary $q$ in (\ref{eq5}) and check the scaling, it is crucial to obtain simulation data for three or more fixed box sizes, say $L_{i}=L_{1},L_{2},\cdots,L_{n}$ $(n\geq 3)$, at the same temperatures $T_{j}$ $(j=0,1,2,\cdots)$.

It must also be stressed that high quality, precise data are essential. These can be obtained, as previous studies of the HCSW fluid \cite{ork:fis:pan} and the RPM \cite{lui:fis:pan} demonstrate, by careful simulation and the use of multiple histogram reweighting \cite{fer:swe}.

More formally, the initial step is to collect grand canonical Monte Carlo data sets for the $Q$-minima, $\{Q_{\mbox{\scriptsize m}}^{\pm}(T,L),\rho_{\mbox{\scriptsize m}}^{\pm}(T,L)\}$, generated at a sufficiently low $T_{0}$ as is to be verified by the ease of fitting to (\ref{eq5}). This is illustrated in Fig.\ 2(a) using data for the HCSW fluid;
\begin{figure}[ht]
\vspace{-0.8in}
\centerline{\epsfig{figure=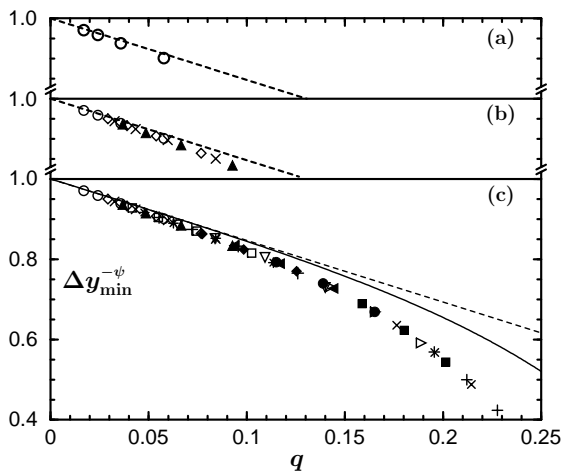,width=3.2in,angle=0}}
\vspace{-0.9in}
\caption{Scaling plot of $\Delta y_{\mbox{\scriptsize min}}^{-\psi}$ vs.\ $q\equiv \bar{Q}_{\mbox{\scriptsize min}}\ln (4/e\bar{Q}_{\mbox{\scriptsize min}})$ with $1/\psi=0.326$ for the HCSW fluid; (a) at $T_{0}\simeq 0.903\, T_{\mbox{\scriptsize c}}$, (b) at $T_{0}$ and three higher temperatures up to $T\simeq 0.952\, T_{\mbox{\scriptsize c}}$, and (c) up to $T\simeq 0.985 T_{\mbox{\scriptsize c}}$. The dashed lines represent the exact two-Gaussian limiting form to linear order in $q$; the solid curve in (c) portrays the full two-Gaussian approximation which evidently deviates significantly from the HCSW results even for $q\simeq 0.05$.\label{fig2}}
\end{figure}
 but note, in particular, that the magnitude of the (positive) exponent $\psi$ is arbitrary and may be assigned any graphically convenient value (such as, e.g., $\psi=2$ or $5$: see \cite{kim:fis:lui}). However, the reason for the choice made will be explained below.

To generate the scaling function successfully, $n=3$ distinct box sizes may well suffice although $n=4$ has been used in our calculations. Furthermore, in order to avoid effects arising from {\em corrections} to the leading scaling behavior, the $L_{i}^{\ast}\equiv L_{i}/a$ should not be too small (where $a$ measures the particle size). For the HCSW fluid $L^{\ast}\gtrsim 7$ sufficed.

At an appropriate $T_{0}$ the value of $\bar{Q}_{\mbox{\scriptsize min}}$ will be small: for our choice of $T_{0}$ for the HCSW fluid we had $\bar{Q}_{\mbox{\scriptsize min}}\lesssim 0.03$; but a somewhat larger value might provide acceptable accuracy. Following the prescription, one density-discontinuity value, say $\Delta\rho_{T_{0}}$ is then chosen for all the $L_{i}$ to provide the best fit of $\Delta y_{\mbox{\scriptsize min}}^{(i)}\equiv [\rho_{\mbox{\scriptsize m}}^{+}(T_{0},L_{i})-\rho_{\mbox{\scriptsize m}}^{-}(T_{0},L_{i})]/\Delta\rho_{T_{0}}$ to the relation (\ref{eq5}) with $q_{0}^{(i)}\equiv q(T_{0},L_{i})$. For the HCSW fluid this fit could be achieved accurately to within $\pm 0.2\%$ of $\Delta\rho_{T_{0}}$. One may also check that the assignment of $\psi$ in any reasonable range has negligible effect on the fitting (which should be weighted more heavily on the smaller values of $q_{0}^{(i)}$). The optimal value $\Delta\rho_{T_{0}}$ is then identified as an estimate for $\Delta\rho_{\infty}(T_{0})$.

The next step is to increase $T_{0}$ to $T_{1}=T_{0}+\Delta T_{0}$ and  to compute the new data sets $\{ \Delta y_{\mbox{\scriptsize min}}(T_{1};L_{i})\}_{i=1}^{n}$ and $\{ q_{1}^{(i)}\equiv q(T_{1};L_{i})\}_{i=1}^{n}$. In doing so, however, the crucial point is to select $\Delta T_{0}$ small enough that the new set $\{ q_{1}^{(i)}\}_{i=1}^{n}$ {\em overlaps} the previous one $\{ q_{0}^{(i)}\}_{i=1}^{n}$. When the new data set is in place, one must find, as before, a new value, $\Delta\rho_{T_{1}}$, such that the new data when plotted overlap and smoothly extend the previous data: see Fig.\ 2(b). The procedure thereby extends and numerically validates the scaling function up to larger values of $q$. The new value $\Delta\rho_{T_{1}}$ can then be taken as an optimal estimate for $\Delta\rho_{\infty}(T_{1})$. 

Subsequently, repeating these steps by increasing the temperature to $T_{j+1}=T_{j}+\Delta T_{j}$ will extend the scaling function further and generate successive estimates $\Delta\rho_{\infty}(T_{j})$ for $j=2,3,\cdots$: see Fig.\ 2(c). As $T_{\mbox{\scriptsize c}}$ is approached, one will observe that smaller increments $\Delta T_{j}$ are needed and high quality data become increasingly important.

Since, via (\ref{eq0}), $\Delta\rho_{\infty}(T)$$\,\rightarrow\,$$0$ when $T$$\,\rightarrow\,$$T_{\mbox{\scriptsize c}}-$ whereas the interval $\rho_{\mbox{\scriptsize m}}^{+}(T;L)-\rho_{\mbox{\scriptsize m}}^{-}(T;L)$ does not then vanish, the plot of $\Delta y_{\mbox{\scriptsize min}}^{-\psi}(q)$ must approach zero at criticality. This behavior is clear in Fig.\ 3 which presents the full scaling function,
\begin{figure}[ht]
\vspace{-0.9in}
\centerline{\epsfig{figure=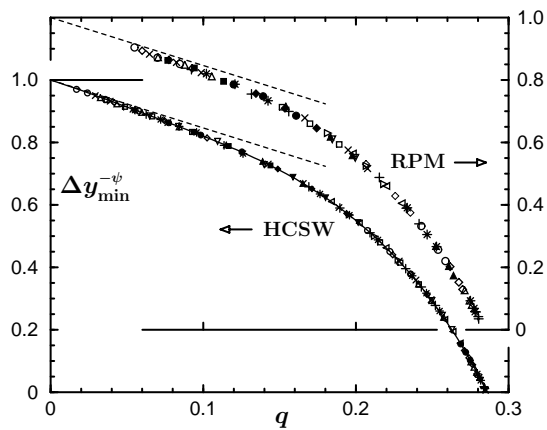,width=3.2in,angle=0}}
\vspace{-0.9in}
\caption{Scaling plots of $\Delta y_{\mbox{\scriptsize min}}^{-\psi}$ vs.\ $q$ for the HCSW fluid and the RPM with $1/\psi$ identified with the Ising exponent $\beta\simeq 0.326$. The dashed lines again show the exact two-Gaussian limit to linear order while the solid curve (passing through the HCSW data) represents the approximant (\ref{eq11}).\label{fig3}}
\end{figure}
 $[\mbox{\boldmath $\mathcal Y$}(q)]^{-\psi}$, as constructed via the algorithm both for the HCSW fluid and for the RPM.

The vanishing of $\Delta y_{\mbox{\scriptsize min}}^{-\psi}$ at $q=q_{\mbox{\scriptsize c}}\simeq 0.2860$ generates unequivocal estimates for $T_{\mbox{\scriptsize c}}$. For the HCSW fluid, a precision of $\pm 2$ parts in $10^{5}$ is realized: furthermore, the value for $T_{\mbox{\scriptsize c}}$ agrees well with less precise ($\pm 3$ parts in $10^{4}$) estimates obtained by studying the model only {\em above} $T_{\mbox{\scriptsize c}}$ \cite{ork:fis:pan}.

\begin{figure}[ht]
\vspace{-0.9in}
\centerline{\epsfig{figure=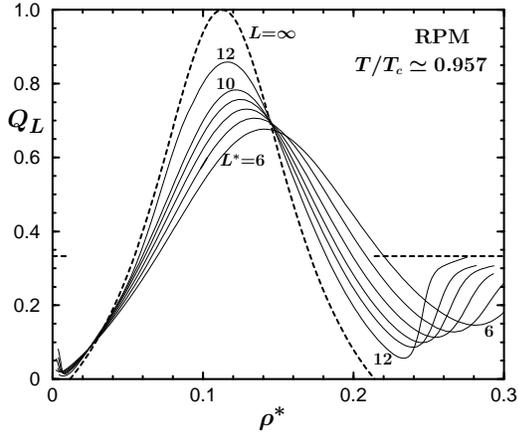,width=3.2in,angle=0}}
\vspace{-1.0in}
\caption{Simulation data for the restricted primitive model (RPM) electrolyte as in Fig.\ \ref{fig1}; but notice the significantly greater asymmetry. \label{fig4}}
\end{figure}

The $Q_{L}(T;\langle\rho\rangle_{L})$ data for the RPM are harder to generate accurately and, moreover, as seen in Fig.\ \ref{fig4},
 the variation of $Q_{L}$ with $\langle\rho\rangle_{L}$ turns out to be highly asymmetric, in strong contrast to the HCSW data in Fig.\ \ref{fig1}. (It might be remarked that the marked asymmetry of the RPM seems to be associated with a large, ${\mathcal R}_{\mu}\simeq 0.26$, value of the Yang-Yang ratio \cite{kim:fis:lui}.) Nevertheless, the algorithm continues to work surprisingly well, yielding $\Delta\rho_{\infty}(T)$ to within $\pm 1$-$2\%$ down to $|t|\simeq 10^{-3}$ and generating estimates for $T_{\mbox{\scriptsize c}}$ with a precision of 4 parts in $10^{4}$: these in turn agree completely with previous, $T>T_{\mbox{\scriptsize c}}$ estimates \cite{lui:fis:pan}. It is important to note, furthermore, that to within the uncertainties, the RPM data in Fig.\ \ref{fig3} fully confirm the expected universality  of the scaling function {\boldmath $\mathcal Y$}$(q)$.

\begin{figure}[ht]
\vspace{-0.9in}
\centerline{\epsfig{figure=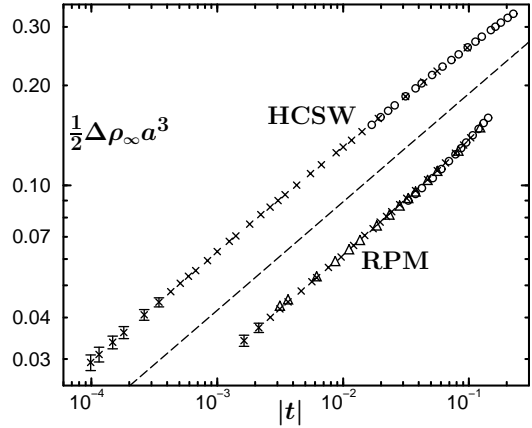,width=3.3in,angle=0}}
\vspace{-1.0in}
\caption{A log-log plot of the coexistence-curve half-width, $\frac{1}{2}\Delta\rho_{\infty}a^{3}$ vs.\ $t\equiv (T-T_{\mbox{\scriptsize c}})/T_{\mbox{\scriptsize c}}$ (where $a$ is the hard-sphere diameter) for the HCSW fluid and for the RPM (at a $\zeta=5$ fine-discretization level \cite{kim:fis:lui,lui:fis:pan}). The crosses follow from the full unbiased scaling algorithm while the circles for $|t|$$\,>\,$$10^{-2}$ report the best previous estimates (employing an equal-weight prescription). The triangles for the RPM are estimates obtained from the simple, economical (but biased) algorithm embodied in (\ref{eq11a}) and (\ref{eq11}). The dashed line has a slope corresponding to $\beta=\beta_{\mbox{\scriptsize Is}}\simeq 0.326$: see (\ref{eq0}).\label{fig5}}
\end{figure}

The analysis presented above demonstrates that $\Delta y_{\mbox{\scriptsize min}}^{-\psi}$ must vanish {\em linearly} with $q$ if one sets $1/\psi =\beta$; but note again that this choice is {\em not} needed in order to estimate $T_{\mbox{\scriptsize c}}$ reliably. However, the HCSW fluid is certainly expected to exhibit Ising-type criticality with $\beta_{\mbox{\scriptsize Is}}\simeq 0.326$; and this is convincingly confirmed by the resulting plot of the coexistence curve shown in Fig.\ \ref{fig5}.
 Accordingly, $\psi$$\,=\,$$1/\beta_{\mbox{\scriptsize Is}}$ was selected for use in Figs.\ \ref{fig2} and \ref{fig3}. A bonus of the RPM calculations also displayed in Fig.\ \ref{fig5} is that $\beta$ close to $\beta_{\mbox{\scriptsize Is}}$ again fits well. This result is of value since serious doubts have been raised regarding the universality class of ionic systems \cite{kim:fis:lui,lui:fis:pan,wei:sch,kim:fis4}.

\section{An Economical Biased Algorithm}
\label{sec4}

In practice, the full, unbiased scaling algorithm may be inconvenient for some applications since it requires a significant amount of precise data narrowly spaced in temperature, especially when $T_{\mbox{\scriptsize c}}$ is approached. Furthermore, one needs to calculate reliably an initial set of $Q$ minima at sufficiently low $T$ that the two-Gaussian structure of ${\mathcal P}_{L}(\rho;T)$ is well obeyed. On the other hand, if one is prepared to accept that a model of interest exhibits Ising-type criticality, one can, in fact, utilize the HCSW scaling plot for $y_{\mbox{\scriptsize min}}(T;L)$ in Fig.\ \ref{fig3} to estimate $\Delta\rho_{\infty}$ at {\em any} given $T$! 

To see this most clearly, recall that $\Delta y_{\mbox{\scriptsize min}}(T;L)$ is (for large enough $L$ and, say, $|t|$$\,=\,$$(T_{\mbox{\scriptsize c}}-T)/T_{\mbox{\scriptsize c}}$$\,\lesssim\,$$0.1$) described by a universal scaling function, {\boldmath $\mathcal Y$}$(q)$, as Fig.\ \ref{fig3} demonstrates. Then we may rewrite (\ref{eq4}) in the direct form
 \begin{equation}
  \Delta\rho_{\infty}(T) \approx [\rho_{\mbox{\scriptsize m}}^{+}(T;L)-\rho_{\mbox{\scriptsize m}}^{-}(T;L)]/\mbox{\boldmath $\mathcal Y$}[q(T;L)],  \label{eq11a}
 \end{equation}
where $q(T;L)$ follows from (\ref{eq3}) and (\ref{eq10}). In words this simply says that {\boldmath $\mathcal Y$}$(q)$ acts as a {\em correction factor} that transforms the first approximation to the coexistence-curve width derived from the locations of the $Q$-minima, into the desired answer. Thus, for a selected temperature $T$ one need only locate the minima of $Q_{L}(T;\langle\rho\rangle_{L})$ (for a reasonable value of $L$), determine the difference $(\rho_{\mbox{\scriptsize m}}^{+}-\rho_{\mbox{\scriptsize m}}^{-})$, calculate $q$, and substitute in (\ref{eq11a}). As a wise precaution, using a second value of $L$ will enable one to check that corrections to scaling are negligible.

To facilitate this very simple, albeit biased procedure, we have fitted the HCSW data in Fig.\ \ref{fig3} to an expression for {\boldmath $\mathcal Y$}$(q)$ that, with $\beta=\beta_{\mbox{\scriptsize Is}}\simeq 0.326$, embodies the linear vanishing when $q\rightarrow q_{\mbox{\scriptsize c}}\simeq 0.2860$ and the exact small-$q$ behavior (\ref{eq5}). Indeed, with $\tilde{q}\equiv q/q_{\mbox{\scriptsize c}}$, the approximant
 \begin{eqnarray}
 & &[\mbox{\boldmath $\mathcal Y$}(q)]^{-1/\beta} \nonumber \\
  &  & \hspace{0.2in}\simeq \left(1-\frac{q}{2\beta}\right)\frac{(1-\tilde{q})(1+a_{2}\tilde{q}^{2}+a_{3}\tilde{q}^{3})}{1-\tilde{q}+b_{2}\tilde{q}^{2}+b_{3}]\tilde{q}^{3}},  \label{eq11}
 \end{eqnarray}
provides an excellent fit (see Fig.\ 3) for the coefficient values $a_{2}=1.829$, $a_{3}=1.955$, $b_{2}=2.340$, and $b_{3}=-1.388$.

We have tested this approach on the RPM (where Ising-type criticality is now well established \cite{lui:fis:pan,kim:fis4}): it yields the triangular data points shown in Fig.\ \ref{fig5}. These evidently agree well with the results of the full, unbiased algorithm. Thus we believe that the approximant (\ref{eq11}) provides a convenient practical tool for accurately estimating the coexistence curves for a wide range of systems of Ising character.
\vspace{-0.3in}
\section{Summary}
\label{sec5}

In conclusion we have presented a novel scaling algorithm developed to estimate precisely the coexistence curves of asymmetric fluids near criticality. Both the fuller unbiased and a simpler biased approach have been illustrated using simulation data for the HCSW fluid and the RPM: precise and reliable estimates for $\Delta\rho_{\infty}$$\,\equiv\,$$\rho^{+}(T)-\rho^{-}(T)$ can be obtained even very close to $T_{\mbox{\scriptsize c}}$. The biased approach, using (\ref{eq11a}) and the accurate scaling function representation (\ref{eq11}), should prove especially useful in exploratory investigations. On the other hand, the full algorithm extends to yield estimates of the coexistence diameter, (\ref{eq.diam}), and the Yang-Yang ratio \cite{kim:fis:lui,fis:ork}.

The support of the National Science Foundation (through Grant No.\ CHE 03-01101) is gratefully acknowledged. Jean-No\"{e}l Aqua and Sarvin Moghaddam kindly commented on a draft.


\begin{thebibliography}{00}
 \bibitem{pan} See, e.g., A.\ Z.\ Panagiotopoulos, Mol.\ Simul.\ 9 (1992) 1.
 \bibitem{kim:fis:lui} Y.\ C.\ Kim, M.\ E.\ Fisher, E.\ Luijten, Phys.\ Rev.\ Lett.\ 91 (2003) 065701.
 \bibitem{bor:kap} C.\ Borgs, S.\ Kappler, Phys.\ Lett.\ A 171 (1992) 37.
 \bibitem{ork:fis:pan} G.\ Orkoulas, M.\ E.\ Fisher, A.\ Z.\ Panagiotopoulos, Phys.\ Rev.\ E 63 (2001) 051507.
 \bibitem{bin} K.\ Binder, Z.\ Phys.\ B 43 (1981) 119.
 \bibitem{bin:lan} K.\ Binder, D.\ P.\ Landau, Phys.\ Rev.\ B 30 (1984) 1477; B.\ D\"{u}nweg, D.\ P.\ Landau, Phys.\ Rev.\ B 48 (1993) 14182.
 \bibitem{bru:wil} A.\ D.\ Bruce, N.\ Wilding, Phys.\ Rev.\ Lett.\ 68 (1992) 193.
 \bibitem{kim:fis} Y.\ C.\ Kim, M.\ E.\ Fisher, J.\ Phys.\ Chem.\ B 108 (2004) 6750.
 \bibitem{kim:fis2} Y.\ C.\ Kim, M.\ E.\ Fisher, to be published.
 \bibitem{fis:ork} M.\ E.\ Fisher, G.\ Orkoulas, Phys.\ Rev.\ Lett.\ 85 (2000) 696 .
 \bibitem{fis:bar} M.\ E.\ Fisher, M.\ N.\ Barber, Phys.\ Rev.\ Lett.\ 28 (1972) 1516.
 \bibitem{kim:fis3} Y.\ C.\ Kim, M.\ E.\ Fisher, Phys.\ Rev.\ E 68 (2003) 041506.
 \bibitem{lui:fis:pan} E.\ Luijten, M.\ E.\ Fisher, A.\ Z.\ Panagiotopoulos, Phys.\ Rev.\ Lett.\ 88 (2002) 185701.
 \bibitem{fer:swe} A.\ M.\ Ferrenberg, R.\ H.\ Swendsen, Phys.\ Rev.\ Lett.\ 63 (2002) 1195.
 \bibitem{wei:sch} H.\ Weing\"{a}rtner, W.\ Schr\"{o}er, Adv.\ Chem.\ Phys.\ 116 (2001) 1.
 \bibitem{kim:fis4} Y.\ C.\ Kim, M.\ E.\ Fisher, Phys.\ Rev.\ Lett.\ 92 (2004) 185703.

\end{thebibliography}

\end{document}